# Design of a Conduction-cooled 4T Superconducting Racetrack for a Multi-field Coupling Measurement System

CHEN Yu-Quan(陈玉泉)[1,2;1)] MA Li-Zhen(马力祯)[1] WU-Wei(吴巍)[1] GUAN Ming-Zhi(关明智)[1] WU Bei-Min(吴北民)[1] MEI En-Ming(梅恩铭)[1] Xin Can-Jie(辛灿杰)[3]

[1] Institute of Modern Physics, Chinese Academy of Sciences, Lanzhou 730000, China
[2] University of Chinese Academy of Sciences, Beijing 100049, China
[3] Key Laboratory of Mechanics on Environment and Disaster in Western China, The Ministry of Education of China, College of Civil Engineering and Mechanics, Lanzhou University, Lanzhou, 730000, P.R. China

*Abstract*—A conduction-cooled superconducting magnet producing a transverse field of 4 T has been designed for a new generation multi-field coupling measurement system, which will be used to study the mechanical behavior of superconducting samples at cryogenic temperatures and intense magnetic fields. A compact cryostat with a two-stage GM cryocooler designed and manufactured for the superconducting magnet. The magnet is composed of a pair of flat racetrack coils wound by NbTi/Cu superconducting composite wires, a copper and stainless steel combinational former and two $Bi_2Sr_2CaCu_2O_y$ superconducting current leads. The two coils are connected in series and can be powered with a single power supply. In order to support the high stress and attain uniform thermal distribution in the superconducting magnet, a detailed finite element (FE) analysis has been performed. The results indicate that in operating status the designed magnet system can sufficiently bear the electromagnetic forces and has a uniform temperature distribution.

*Index Terms*—conduction-cooled, superconducting magnet, multi-field coupling measurement system, mechanical and thermal analysis

## 1 Introduction

With the diverse applications of superconductors, the knowledge of their mechanical properties in complicated fields is indispensable [1]. However, since superconducting composites are usually sensitive to strain, the investigation of multi-deformation behavior under complicated fields is most important. Recently, at Lanzhou University, China, a new generation multi-field coupling measurement system has been built (see Fig.1). The integrated apparatus includes a magnet subsystem, variable cryogenic subsystem, loading subsystem, and deformation measurement subsystem. Different from some similar instruments in other countries [2-5], which are mainly based on engineering detection of material in a fixed field, the coupling measurement system can be an integrated test facility and system of measurements by contact transducers and contactless signals for temperature/magnetic/electric/mechanical properties in variable and complicated fields.

Considering experimental costs, a cryocooler-cooled superconducting magnet is used due to its simplicity, compactness, and efficiency. This technology can allow the user to obtain high magnetic fields in a common laboratory, where the use of liquid helium is difficult or expensive. In the apparatus two GM cryocoolers are applied to cool down the superconducting magnet and samples to 4.2 K.

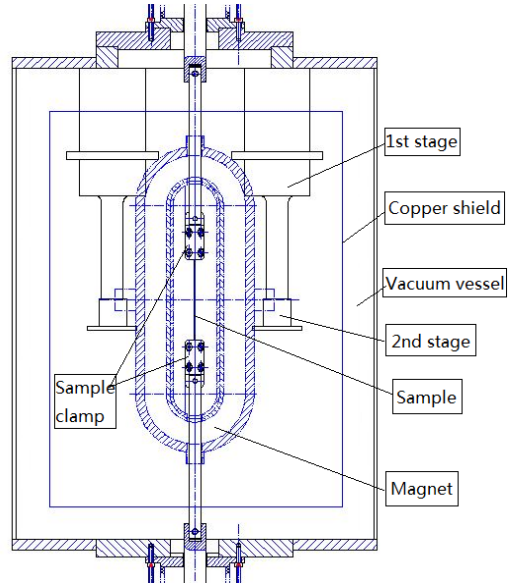

Fig.1. Schematic of the cryocooler-cooled 4T superconducting magnet system and the stretch equipment

## 2 Superconducting magnet system design

### 2.1 Magnetic field design

The overall goal of the superconducting magnet is to produce a magnetic field of 4 T over a length of 150 mm inside a bore of 30 mm diameter. There are several winding types which can generate the transverse fields. To coordinate with the strain measurement system of the nonlinear CCD optics, which needs a gap to observe sample strain by contactless optical signals, a superconducting racetrack type will be used. The magnetic field of two flat racetrack superconducting coils was designed and optimized by using the Opera-TOSCA routine [6]. Fig. 2 shows the coil model created by Opera with the transverse field in the y direction. The parameters of the superconducting racetrack coils are listed in Table 1. The results of the magnetic field simulation are shown in Fig. 3. Fig. 3(a) and Fig. 3(b) show the field in the y direction on a straight line along 150 mm of the z axis and on a cylindrical surface with 10 mm diameter, respectively. It can be seen that the central field reaches 4 T and the field homogeneity has reached the requirement of ±2%.

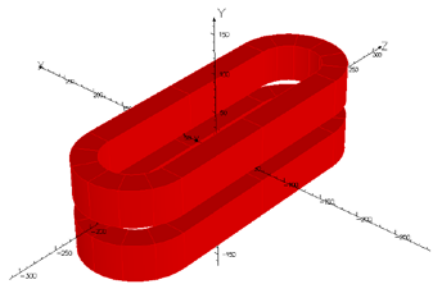

Fig.2. Racetrack type coil model of the design magnet

Table 1. Main parameters of the racetrack coils

| Parameter | Value |
| --- | --- |
| Length of the straight side | 260 mm |
| Radius of the arc side | 39 mm |
| Width in x direction | 30 mm |
| Thickness in y direction | 47 mm |
| Distance between two coils | 24 mm |
| Operating current density | 254 A/mm2 |
| Stored energy | 51 kJ |

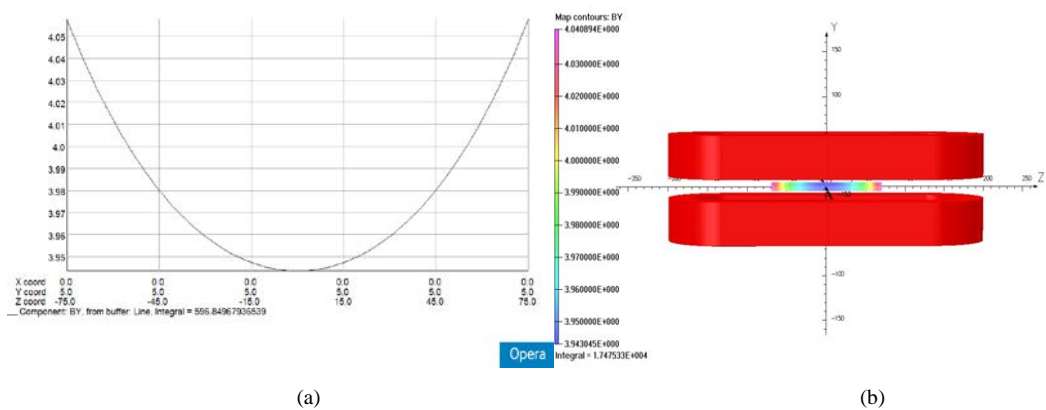

(a)                                                                      (b)

Fig.3. Transverse magnetic field distribution (a) along the central axis and (b) on a cylindrical surface of 10 mm diameter

2.2 Wire selection

We select rectangular cross-section superconducting wires for the racetrack coil winding due to their relatively high filling factor (more than 80%). The superconducting wire has a 1.25 mm×0.8 mm cross section. Each of the racetrack coils is wound with 1260 turns. The magnet stores 51 kJ of energy and on load line the ratio of operating current to critical current is 72%, which is reasonable for a conduction-cooled superconducting magnet.

3 Mechanical and thermal analysis

3.1 Mechanical analysis

In superconducting magnets, where the field and current density are both high, the superconducting wires will sustain a very large Lorentz force F=J×B per unit volume, which can lead to remarkable deformation of the magnet coils. The deformation has potential to lead to the interruption of the superconducting magnet exciting, i.e., quenching. The magnet former must therefore be sturdy enough to support the coils against the large force, so it is important and necessary to calculate the displacements and stresses of the superconducting magnet. In order to obtain enough mechanical strength and heat conduction, 316 stainless steel and copper materials are both used to fabricate the magnet former. Fig. 4 shows the mechanical drawing of the structure. The innermost layer of the former is made of 316 stainless steel material as support, which has yield strength as high as 600MPa at 4.2K. A hollow racetrack type copper block with two grooves clings outside the support. The coils are then wound in the grooves and a pair of aluminum alloy rings is assembled outside so that the coils can be fixed well.

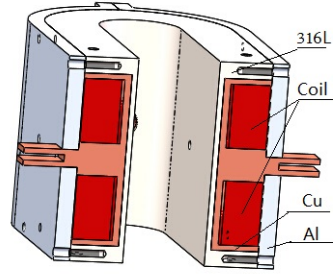

Fig. 4. Sketch of the magnet former and coils

The mechanical analysis of the coils and former were performed by the finite element software ANSYS [7]. In order to simplify the calculation, we have created a one-eighth size model of the magnet system using the element type of solid 186 with 20 nodes. The effects of Lorentz force as well as thermal contraction were considered in this analysis. The method explained in [8] is employed to get the Lorentz forces on each node and then these forces are exerted on the coil model.

In fact, the superconducting coils may be regarded as unidirectional composites from the structural point of view. In this analysis the coil and coil former were regarded as a whole model. In this model, the effective elastic modulus of the superconducting magnet is used to solve the stress problem based on the homogenization method. The average mechanical properties at 4.2K were obtained for the superconducting magnet [9].The displacement and the Von Mises equivalent stress of the coil due to magnetic forces and cool-down are shown in Fig. 5. The maximum displacement is 0.69 mm, which occurs at the curve segment of the coil. The maximum stress is 110 MPa, which is far below the yield stress of 630 MPa of the NbTi/Cu composite wire at 4.2 K [10]. Fig. 6 shows the stresses of the components of the magnet structure. The maximum Von-Mises equivalent stresses on stainless steel former, copper groove and aluminum alloy hoop were, respectively, 394 MPa, 166 MPa and 166 MPa. These stresses are all below the yield strengths of the corresponding materials of each component, so we can consider the magnet former as reasonably designed to support large magnetic forces.

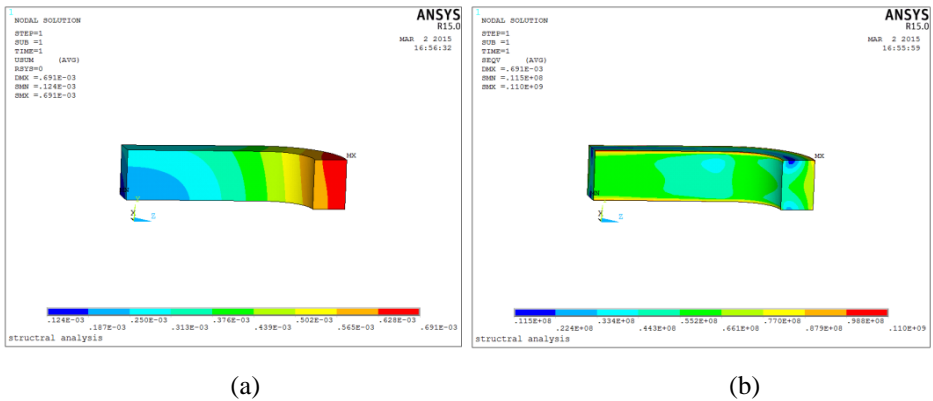

(a)          (b)

Fig.5. The displacement (a) and stress (b) of the coil due to magnetic forces and cool-down

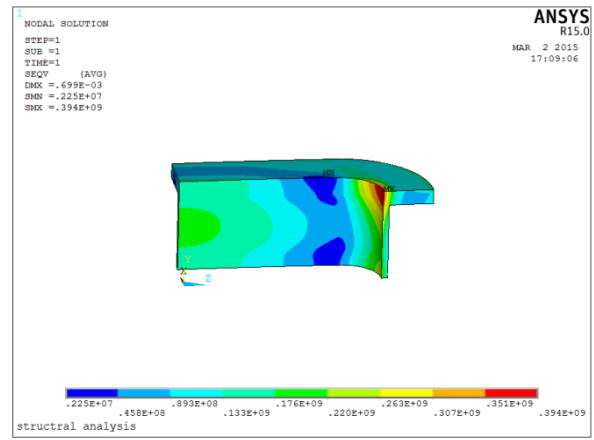

(a)

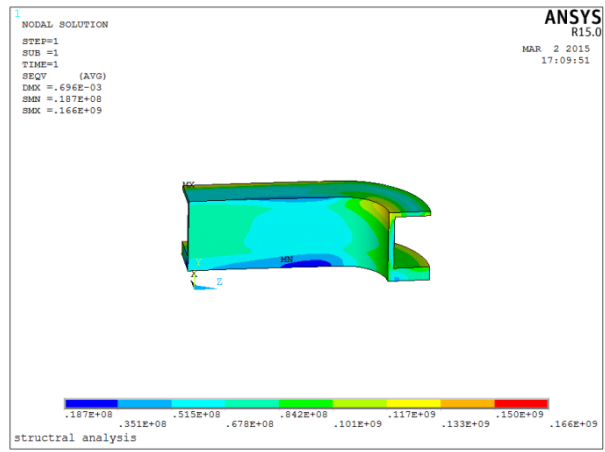

(b)

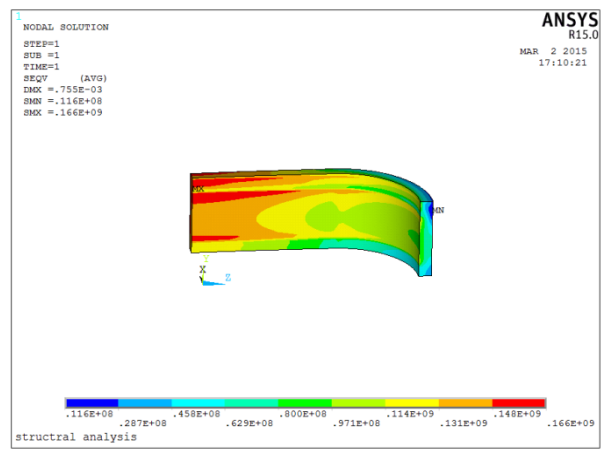

(c)

Fig.6. The stresses of the structure components: (a) 316 stainless steel former, (b) copper groove, (c) aluminum alloy hoop

3.2 Thermal analysis

Unlike liquid helium bath magnets, conduction-cooled magnets are cooled down using cryocoolers. One GM refrigerator, however, can only provide a dynamic cooling capacity of 1.5 W to the 4.2 K reservoir. The system heat losses and thermal distribution of the magnet, which are critical for the stable operation of the magnet, must be calculated. The temperature distribution of the magnet was simulated using ANSYS thermal FEM module [11]. We are only concerned with the temperature on the magnet so the heat loads on the second stage will be calculated. These heat loads include the heat conduction leakage of the HTS current leads, the heat radiation leakage from the copper shield to the magnet and the conduction leakage of the eight epoxy rods (each of diameter 6 mm and length 300 mm from thermal shield to magnet). A pair of HTS current leads produced by Innova Superconductor Technology Co. has a heat leakage of 0.46 W when the temperature ranges from 64 K to 4.2 K. Based on the method presented in [12], the total heat loads on the second stage were estimated at 0.52 W. These heat loads match the operating temperature of the second stage of 4 K. The heat flux on the magnet surface is 0.085 W/m$^2$ when 12 layers of multi-layer insulation are wrapped around the magnet. The total cold mass of the magnet system is about 40 kg.

Static thermal analysis of the structure, which just needs to set the thermal conductivity for the materials, was carried out. The material thermal conductivity at 4.2 K for different parts of the structure is summarized in Table 2. The simulation result is shown in Fig. 8. The results show that the temperature of the magnet is distributed uniformly.

Table 2. Material thermal conductivity at 4.2 K

| Material | Thermal conductivity (W/m/K) |
|---|---|
| Coil | 181 |
| 316 stainless steel | 0.27 |
| Copper | 384 |
| Al6061 | 5.15 |

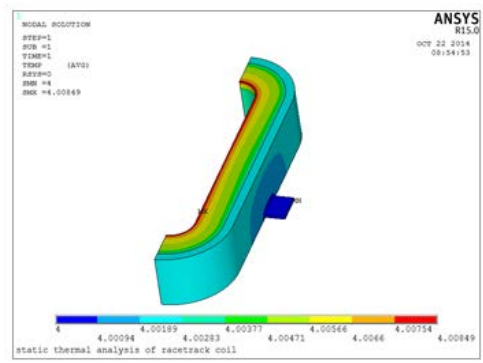

Fig. 7. Temperature distribution of the magnet structure

4 Conclusion

In this paper, the design of a 4 T conduction-cooled superconducting racetrack type magnet was presented. The structural analysis of the coils and former were performed by ANSYS software. Besides the magnetic forces, we also consider the effect of the thermal contraction from 293 K to 4.2 K. The maximum displacement of the coil is 0.69 mm, which occurs at the curve segment of the coil. The maximum stresses on each component of the coil former are all below the yield strengths of the corresponding materials. The temperature distribution of the magnet is simulated using the static thermal model and the temperature of the coils has a uniform distribution. Therefore, it demonstrates that the design is reasonable for a conduction-cooled magnet.